\documentclass[paper]{ieice}
\usepackage[varg]{txfonts}
\usepackage{hyperref}
\usepackage{cite}
\usepackage{setspace}
\usepackage{scalerel}

\usepackage{graphicx}
\graphicspath{{./pdf/}{./jpeg/}{./png/}{./jpg/}}
\DeclareGraphicsExtensions{.pdf,.jpeg,.png,.jpg}
\usepackage{flushend}

\field{}
\title{Integrating UML with Service Refinement for \\Requirements Modeling and Analysis}

\authorlist{
 \authorentry[yylonly@gmail.com]{Yilong Yang}{m}{um}
 \authorentry[yangjing6646@yahoo.com.hk]{Jing Yang}{n}{gz}
 \authorentry[xsl@umac.mo]{Xiaoshan Li}{n}{um}
}
\affiliate[um]{The authors are with the Department of Computer and Information Science, Faculty of Science and Technology, University of Macau} 
\affiliate[gz]{The author is with the College of Computer Science and Information, Guizhou University, China}

\begin{document}
\maketitle 
\begin{summary}
Unified Modeling Language (UML) is the de facto standard for requirements modeling and system design. UML as a visual language can tremendously help customers, project managers, and developers to specify the requirements of a target system. However, UML lacks the ability to specify the requirements precisely such as the contracts of the system operation, and verify the consistency and refinement of the requirements. These disadvantages result in that the potential faults of software are hard to be discovered in the early stage of software development process, and then requiring to pay much efforts in software testing to find the bugs. Service refinement is a formal method, which could be a supplement to enhance the UML. In this paper, we show how to integrate UML with service refinement to specify requirements, and verify the consistency and refinements of the requirements through a case study of online shopping system. Particularly, requirements are modeled through UML diagrams, which includes a) use case diagram, b) system sequence diagrams and c) conceptual class diagram. Service refinement enhances the requirements model by introducing the contracts. Furthermore, the consistency and refinements of requirement model can be verified through service refinement. Our approach demonstrates integrating UML with service refinement can require fewer efforts to achieve the consistency requirements than only using UML for requirement modeling. 
\end{summary}

\section{Introduction}
The methodology of software development has been tremendously evolved from object-oriented, component-based to service-oriented approach \cite{sommerville2015software}. Those methodologies are focus on dealing with the complexity of software \cite{larman2012applying}. UML can help the developer to handle inherent complexity by modelling and analyzing the functionality of the target system. Particularly, the developers can use use-case diagram and system sequence diagrams to specify the requirements model in the procedure of the requirements analysis, which includes defining the services interface and interaction protocols between the interface and environment. Furthermore, the domain model of the target system can be specified by the conceptual class diagram. However, the developers cannot know the potential issue of the requirements only depended on those diagrams, because UML cannot self-verify no bad properties (E.g., dead-lock and live-lock) in requirements models, and it does not contain any techniques to verify the correctness of refinements. That will make an embarrassing situation when several "un-verified" requirements models are specified from the different stage of software development processes, the developers cannot guarantee that the new requirements model is enhanced from the previous one, by adding more the functional requirements without introducing any problems. Those disadvantages are the motivations of this paper to integrate UML with formal methods \cite{saeeiab2000method, oda2017formal, fan2006constraint} to precisely define the requirement model, and verify consistency and refinement of the requirements. 

Service refinement is chosen \cite{hjfo8} to enhance UML, because it is a well-designed formal method on the calculus of contract refinement. It lays the foundation of guard design and design refinement from Unifying Theories of Programming (UTP) \cite{hoare1998unifying}, and the refinement of divergence and failures from Communicating Sequential Processes (CSP) \cite{csp}. It models and verified many large system such as rocket, high-speed train, aerospace system. However, service refinement is not natively designed for object-oriented requirements modeling and analysis. We made some necessary extensions to service refinement before integrating with UML:

\vspace{.1cm}
\noindent 1) \textit{service visibility}. The scope of visibility is used by UML to describe whether an operation of the interface is public or private. Corresponding, we extended the interface of service refinement with visible scope, which can specify whether a service is accessible for the outside environment or only serves as an internal service for other services. 

\vspace{.1cm}
\noindent 2) \textit{divergence by private service}. The divergence state defines the system in the unstable state of livelock, in which the system infinitely invokes internal actions. After extending service refinement with visible scope, divergence state is defined as infinitely invoking the private services of the interface.

\vspace{.1cm}
\noindent 3) \textit{consistency as deadlock-free and livelock-free}. The consistency of service refinement only consider the deadlock. The extended consistency of contract must be deadlock-free and livelock-free. That means after any trace, not all services including private and public services are refusal to the environment, and the trace must not contain the infinitely private service invoking for livelock-free. 

\vspace{.1cm}
\noindent 4) \textit{refinement by hiding private services}. The refinement defined that the interface has the same behavioural with environment, but has less bad properties such as deadlock and livelock. We need hidden the private service to make the refinement definition consistent. Therefore, the extended refinement is defined as hiding the private services from the contract, the refined contract has less divergences and failures.
   
\vspace{.2cm}
\noindent \textbf{Contributions} Online Shopping System (OSS) is the most common used and large system in our daily life. We adopt the case-based approach to present our methods through OSS case study. In summary, our major contributions are:
   
\vspace{.1cm}	
\noindent 1) The extension of service refinement for requirements modeling and analysis by UML.
   
\vspace{.1cm}
\noindent 2) We illustrate that UML can enhance service refinement to elicit start-up requirements, which includes the skeleton of the service interface and domain model through use case diagram and interface and conceptual class diagrams, the draft of protocol and failures of contracts by system sequence diagram.
   
\vspace{.1cm}
\noindent 3) Base on the start-up UML requirements, we demonstrate that service refinement can refine the requirements model more precisely through specifying the contracts and protocols of requirements mathematically.
   
\vspace{.1cm}
\noindent 4) We demonstrate that service refinement can give system analysts the directions of refinement, prove the correctness of refinements and verify the consistency.

\vspace{.1cm}
\noindent 5) To demonstrate our approach works, we use the model checking tool FDR to specify the same contracts of online shopping system, and then verify the deadlock-free and livelock-free as well as the refinement of the contracts.  

\vspace{.3cm}
\noindent The remainder of this paper is organized as follows: Section 2 is preliminary of service refinement and the extensions. Section 3 presents requirement elicitation by UML. And then Section 4 shows formal specifications by service refinement. Section 5 presents refinement of contract and consistency verification. Finally, section 6 concludes this paper and discuss the future work.

\section{Service Refinement: Preliminary and Extension}
For self-contained and completeness of our paper, the brief introduction and extension of the service refinement \cite{hjfo8} are presented in this section.

\subsection{Interface} 

\noindent The interfaces are the access points of the system. An interface is defined as a tuple with a resource declaration sector and a service declaration sector:
\begin{center}
  \label{interface}
  \( I=(\textit{RDec}, \textit{SDec}) \)
  \end{center}
  \textit{RDec} is a set of variables, variable defines as $x:T$, where $T$ stands for the
  type of the variable. \textit{SDec} is a set of the services, the service signature $m(in:U, out:V)$
  declares service $m$ holds variables $in$ of type $U$ and variables
  $out$ of type $V$ as its input and output parameters of the service
  respectively.

\vspace{0.1cm}
\noindent\textbf{Extension for Interface} 
\vspace{0.1cm}

\noindent The prefix signature of service indicates the visibility of the service for the environments. We use the same notation of interface described in UML. The prefix notation "+" indicates the service is a public service. The notation prefix "-" indicates the service is a private service, which is not visible for the environment. The prefix set is $\mathit{prefix = \{+, -\}}$. E.g, a public service $m$ describes as $\mathit{+m(in:U, out:V)}$. In default, the service without a prefix is regarded as a public service.

\vspace{0.1cm}
\noindent\textbf{Semantics of Service} 
\vspace{0.1cm}

\noindent A specification of a service $m$ is a triple $(\alpha_m, g_m, P_m)$, where

\noindent 1) $\alpha_m$ comprises all the resource managed by the service.

\noindent 2) $g_m$ is the firing condition of the service, characterizing the circumstance under the service can be activated.

\noindent 3) $P_m$ is a reactive design, describing the behavior of execution of service.

\subsection{Contract} 

\noindent Contracts are the specifications of interfaces. A contract of an interface specifies the functionality of services declared in the interface, the protocol of the interactions. 

\noindent A contract is a quadruple \(Ctr=(I, Init, \textit{Spec}, Prot)\) where
\vspace{-.2cm}
\begin{description}
     \item [{(1)}] $I$ is an interface.
     \item [{(2)}] $Init$ specifies the initial
          state of the design 
           \begin{center}
           \( Init=true\vdash (init(v')\wedge \neg wait')\)
           \end{center}
           where $v$ stands for resources of interface $I$.
     \item [{(3)}]
           \textit{Spec} maps each service $m(in:U, out:V)$  of interface $I$ to
                   its specification $(\alpha_m, g_m, P_m)$
     \item [{(4)}]$Prot$ is a protocol set of valid traces of service
        activation events, specifying the interaction pattern between the
        contract with its environment, where the event $?m(x)$
        represents the call of service $m$ with the input $x$. The protocol indicates that the contract can provides the normal response if its services are invoked in the orders included in $Prot$. Otherwise, the result will be unpredictable.
\end{description}

\vspace{0.1cm}
\noindent\textbf{Semantics of Contract} 
\vspace{0.1cm}

\noindent The dynamic behaviour of contract $Ctr$ is described by the triple $(\textit{Prot(Ctr)}, \; \textit{Failures(Ctr)}, \; \textit{Divergences(Ctr)})$, where
\vspace{-.1cm}
\begin{itemize}
	\item \textit{Prot(Ctr)} retrieves the weakest protocol of contract $Ctr$.
	\item \textit{Failures(Ctr)} is the set of pairs(s, X) where s is a sequence of interactions between $Ctr$ and its environment, and $X$ denotes a set of events $?m$ and $!m$ in which the contract may refuse to engage after it has performed all actions in s.
    \item \textit{Divergences(Ctr)} consists a set of the sequences of interactions between $Ctr$ with its environment which leads the contract to a divergent state.
\end{itemize} 

\vspace{0.1cm}
\noindent\textbf{Extension of Semantics of Contract} 
\vspace{0.1cm}

\noindent In UML, \textit{divergences(Ctr)} specifies the interactions in an use-case which leads the system into a divergent state, in which a private service $private(m(x))$ is invoked infinitely times. Therefore, the extension of \textit{divergences(Ctr)} is:
   
   	{\small\setlength{\mathindent}{0pt}     
        \[
        \begin{array}{l}
           \mathit{Divergences(Ctr)=_{df}} 
            \mathit{\; \{\langle ?m_{1}(x_{1}), !m_{1}(y_{1}), \dots ,?m_{k}(x_{k}), ?private(m(x)^{*}) \rangle | }\\
            (Init;(g_{m_{1}}\verb|&|P_{m_{1}})[x_{1},y_{1}/x,y'];\dots; \\
            (g_{m_{k-1}}\verb|&|P_{m_{k-1}})[x_{k-1},y_{k-1}/x,y] \mathit{[true,false/ok',wait']}) \\
            (g_{m_{k}}\verb|&|P_{m_{k}})[x_{k}/x] \mathit{[true,false/ok,ok']}) \\ 
        \end{array}
        \]  
    }      

\noindent In UML, the private service can only be invoked from other services in the same interface. Therefore, private services are in the refusal set $X$ of $\mathit{Failures(Ctr)}$, except when the service $m_k$ invokes private service $\mathit{private(m)}$, that means 

{\small\setlength{\mathindent}{0pt}
           \[
           \begin{array}{l}
           	\mathit{Failures(Ctr)=_{df}} \\ \\
           	\left \{
           	\begin{array}{l}
           		 (\langle ?m_{1}(x_{1}), !m_{1}(y_{1}), \dots, ?m_{k-1}(x_{k-1},!m_{k-1}(x_{k-1}))\rangle, \; X) \; | \\
           		\exists v' \cdot (Init;\dots;(g_{m_{k-1}}\verb|&|P_{m_{k-1}})[x_{k-1}/x]) \\
                \verb|[|true,false,true,false/ok,wait,ok',wait']  \wedge \\
           		\forall ?m \in X \cdot \neg g_{m}[v'/v]  \wedge 
           		\mathit{?private(m)} \in X \\
           	\end{array}
           	\right \}  \cup 
           \end{array}
           \]
           \[
           \begin{array}{l}
           	
           	\left \{
           	\begin{array}{l}
           	    (\langle ?m_{1}(x_{1}), !m_{1}(y_{1}), \dots, ?m_{k}(x_{k}))\rangle, \; X) \; | \\
                \exists v' \cdot (Init;\dots;(g_{m_{k}}\verb|&|P_{m_{k}})[x_{k}/x]) \\
                \verb|[|true,false,true,true/ok,wait,ok',wait']  \wedge \\
                \forall ?m \in X \cdot \neg g_{m}[v'/v]  \wedge 
                \mathit{?private(m)} \not \in X
           	\end{array}
           	\right \}  
           \end{array}
           \]
}

\subsection{Contract Consistency} 
A contract $Ctr$ is consistent, if it will never enter deadlock states unless its environment violates the  protocol, i.e.,

{
\setlength{\mathindent}{0pt}
\[
  \forall tr \in Prot \cdot \left(
  \begin{array}{l}
     \exists s \in \textit{traces(Ctr)} \cdot s \downarrow \{?\} = tr \wedge \\
     \forall(s, X) \in \textit{failures(Ctr)} \cdot s \downarrow \{?\} \preceq tr \Rightarrow \\ 
     X \not = \{ ?m,!m \; | \; m \in \textit{MDec}\}
   \end{array}
   \right )
\]} 
\noindent where $\mathit{traces(Ctr) =_{df} \{ s \;| \; \exists X \cdot (s,X) \in failures(Ctr)\}}$ and $s \downarrow \{?\}$ represents the subsequences of $s$ which is formed by the input events. The notation $ t \preceq s$ means sequence $t$ is the prefix of $s$.

\vspace{0.1cm}
\noindent\textbf{Extension of Contract Consistency} 
\vspace{0.1cm}

\noindent In the extension, contract is consistent must be both deadlock-free and livelock-free. That means only if after any trace, the refusal set $X$ dose not contain all the private and public services,

\[ 
\mathit{ X \not = \{ ?m,!m \; | \; m \in \textit{public(MDec)} \cap \textit{private(MDec)}\} }
\] 

\noindent and the subsequences trace of $s$  does not include infinitely invoking the private services. 
\[ 
s \downarrow \{?\} \not = \{ \langle ?m^* \rangle \; | \; ?m \in \textit{private(MDec)\}} 
\]

\subsection{Contract Refinement} 
Let \(\mathit{Ctr=(I_{i}, Init_{i}, Spec_{i}, Prot_{i}}) \; (i=1,2)\) are two
      contracts with the same set of services. $Ctr_{1}$ is
      refined by $Ctr_{2}$, denoted by \(Ctr_{1}\sqsubseteq Ctr_{2}
      \), if
      \begin{description}
        \item [{(1)}] \( \textbf{Divergences}(Ctr_{1})\supseteq
        \textbf{Divergences}(Ctr_{2})\)
        \item [{(2)}] \( \textbf{Failures}(Ctr_{1})\supseteq
        \textbf{Failures}(Ctr_{2})\)
      \end{description}

\vspace{0.1cm}
\noindent\textbf{Extension of Contract Refinement} 
\vspace{0.1cm}

\noindent In UML, the trace of divergence and failures contain the private services invoking. We need hidden them to keep the refinement definition consistent.  Let \(\mathit{Ctr=(I_{i}, Init_{i}, Spec_{i}, Prot_{i}}) \; (i=1,2)\) are two contracts with the same set of public services. $PM_1$ and $PM_2$ re private service sets of $Ctr_1$ and $Ctr_2$. $Ctr_{1}$ is refined by $Ctr_{2}$, denoted by \(Ctr_{1}\sqsubseteq Ctr_{2}\), if
      \begin{description}
        \item [{(1)}] \( \textbf{Divergences}(Ctr_{1} \verb|\| PM_1)\supseteq
        \textbf{Divergences}(Ctr_{2}  \verb|\| PM_2)\)
        \item [{(2)}] \( \textbf{Failures}(Ctr_{1} \verb|\| PM_1)\supseteq
        \textbf{Failures}(Ctr_{2} \verb|\| PM_2)\)
      \end{description}
Note that $Ctr \verb|\| PM$ means hidden the private services in $PM$ from $Ctr$. 

Above is the brief introduction of service refinement. In the next section, we will show how service refinement support for requirements analysis with UML.
\section{Requirements Modeling}
Online Shopping System \cite{6960112,hopson2009online} is a form of electronic commerce system which allows consumers to directly buy products from sellers over the Internet. Consumers can search interested products through the website, which displays the same product's availability and pricing at different e-retailers. In recent years, customers can shop online using a range of different computers and devices, including desktop computers, laptops, tablet computers and smart phones. The Alibaba, Amazon, and eBay are the largest companies providing online shopping services for billions of people all over the world.

\subsection{Use Case Diagram} 
\begin{figure}[!htb]
\begin{center}
  \includegraphics[width=0.45\textwidth]{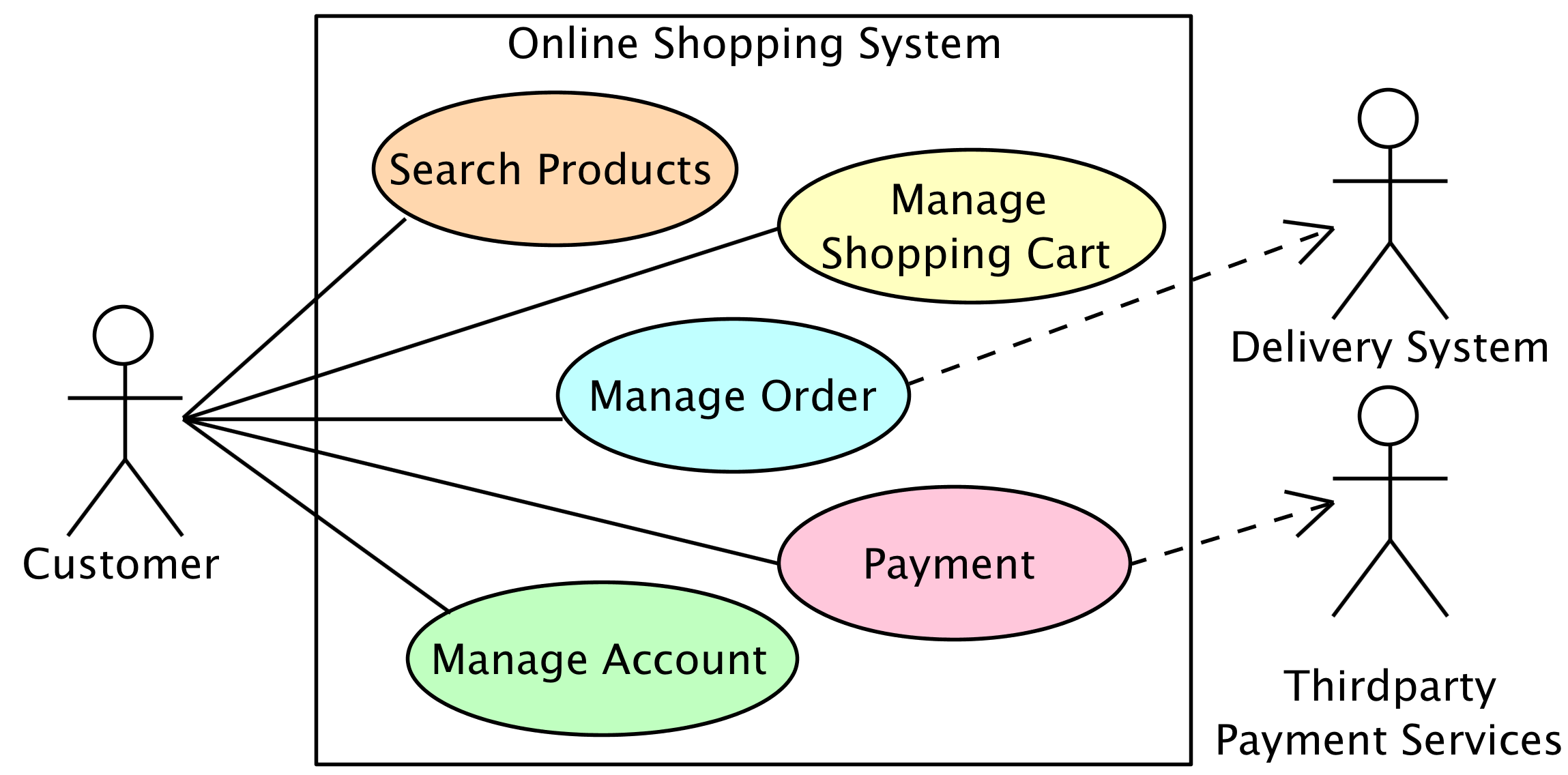}
  \caption{UseCase Diagram on Online Shopping System}
\label{usecase}
\end{center}
\end{figure}   

UML provides use case diagram to help customers specify the requirements of target system. The basic scenarios of shopping online system are specified in Fig. \ref{usecase}:

\vspace{.1cm}
\noindent \textbf{a) Search Products:} The customer can search the desired products by the keywords, and the system displays the candidate products to the customers for further examinations.
  
\vspace{.1cm}
\noindent \textbf{b) Shopping Cart:} When customers found the desired products, they can add the products to the shopping cart if the products are available, e.g., in the stock. The customer can also check the state of the shopping cart in any time, while they find some dislike products in the cart, they can quickly remove them, and when they like to buy more, the customer also can change the number of the item in the shopping car.  E.g., there is already one Nike Shoes in Jack's shopping car, and Jack likes to buy one more for his daddy, he could quickly change the number of shoes in the shopping cart.

\vspace{.1cm}
\noindent \textbf{c) Manage Order:} When shopping cart is ready, the customer can place the order under his account. That means the customer must have an account and log into the system. Moreover, at least one address must be added to their accounts for receiving products. But if the customers have more than one addresses, the system will ask them to choose one for the shipment. Once customers logged into the system, they can check the state of order at any time, which will show whether the order is paid or not. Even the customer can track their products through the delivery system once the their products are sent out. 

\vspace{.1cm}
\noindent \textbf{d) Payment :} While placing the order, the customers must choose the payment approach. If they choose cash on delivery, the cash must be given to the delivery man when receiving products. If the customers choose online payment, the system will check his balance under the account, if the account has not enough balance to pay, the system must ask the customers to pay with their credit cards through third party payment services. 

\vspace{.1cm}
\noindent \textbf{e) Manage Account:}  If customers are the first time to use this online shopping system, they must register membership before checking order out. The customers can modify their information at any time, which includes delivery addresses as well as their credit cards information. 

\subsection{System Interfaces and Domain Concepts Modeling} 
The primary use cases are presented in the previous section. In this section, we use the interface of class diagram of UML to specify the services of use cases in Fig.\ref{interface}.

\begin{figure}[!htb]
\begin{center}
  \includegraphics[width=0.45\textwidth]{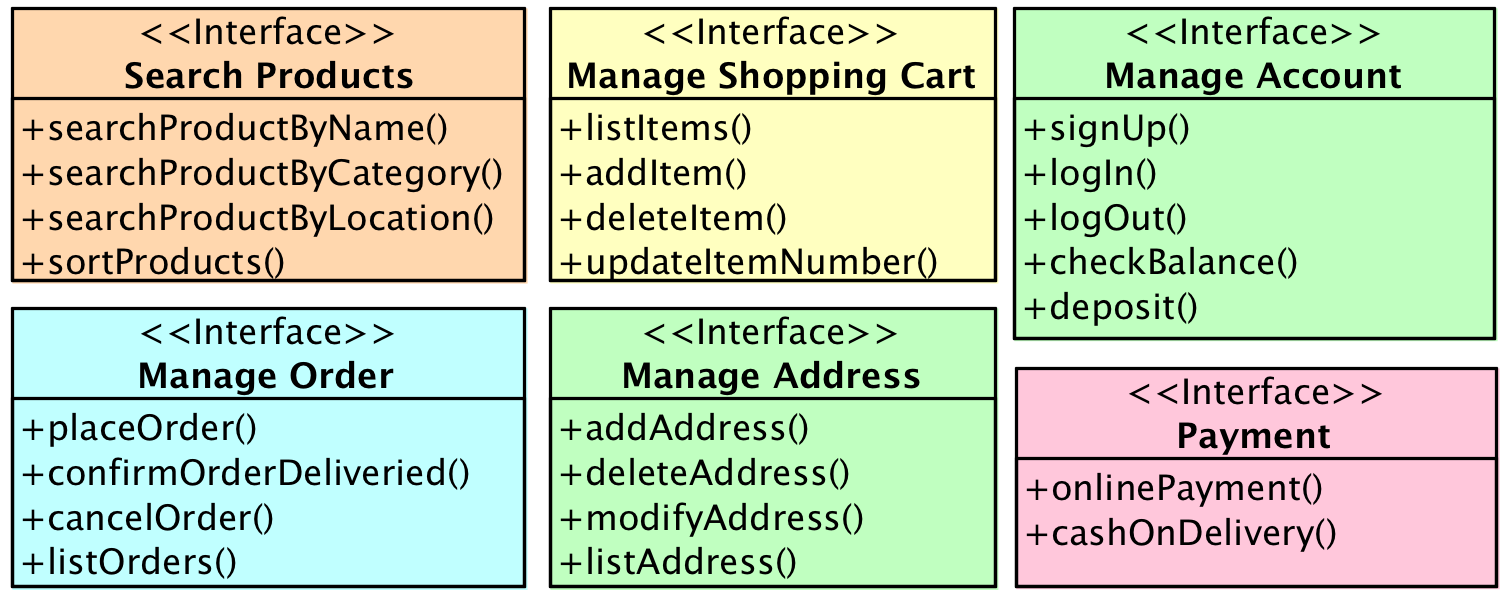}
  \caption{Interfaces of Online Shopping System}
\label{interface}
\end{center}
\end{figure}

For example, the interface of \textit{ManageAccount} contains the service \textit{signup()} for the new customer to register membership, the services \textit{login()} and \textit{logout()} for the member to log into and out the system, \textit{editPersonalInformation()} is used for editing the personal information in this system, \textit{checkBalance()} is used for checking balance in your account and \textit{deposit()} is used for adding the account balance. 

The interface can define all the services of the system, and it even can specify the input and output parameters of interface and required resource. However, UML lacks the ability to define the contract for each service precisely. The formal method, service refinement, not only provides the approach can precisely define service specification, interface, and contract, but also can verify the consistency and the refinement of the contracts. For example, the interface \textit{ManageAccount} can be described by the interface of service refinement as follows:

\vspace{-.2cm}
{\small\setlength{\mathindent}{0pt}
    \[
    \begin{array}{l}
    
         \mathit{ ManageAccountIF \stackrel {\mathrm{def}}{=} (RDec, SDec) } \\ \\
         
           \mathit{SDec} = \{\\
           \quad \mathit{signUp(inPersonalInformation : Customer, \; outR : Boolean)}, \\
           \quad \mathit{logIn(inUserName : String, \; inPasswd : String, \; outR : Boolean)}, \\
           \quad \mathit{logOut(outR : Boolean)}, \\
           \quad \mathit{checkBalance(outBalance : Double)}, \\
           \quad \mathit{deposit(inNewBalance : Double, \; outR : Boolean)} \; \} \\ \\
           
           \mathit{RDec} = \{ \\
          \quad \mathit{UserDB : Customer^*}, \\
          \quad \mathit{CurrentCustomer : Customer}, \\
          \quad \mathit{TempCustomer : Customer}, \\
          \quad \mathit{LoginState : Boolean}\} \\
    \end{array}
    \]}
 \vspace{-.2cm}
 
\begin{figure}[!htb]
\begin{center}
  \includegraphics[width=0.45\textwidth]{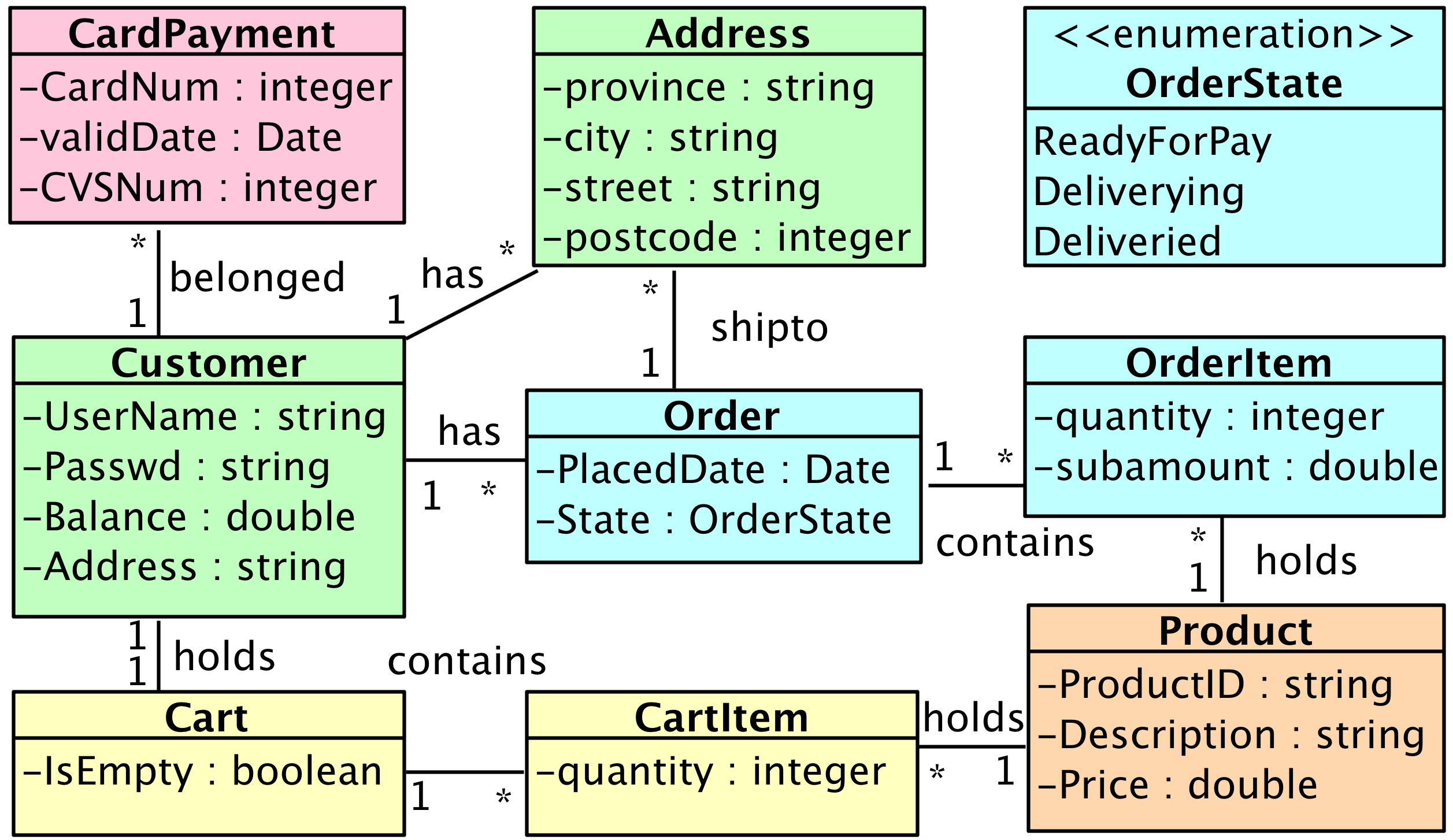}
  \caption{Conceptual Class Diagram of Online Shopping System}
\label{domain}
\end{center}
\end{figure}

\noindent The signatures of services are defined in \textit{SDec}. For example, service \textit{login()} requires variables \textit{username} and \textit{password} typed \textit{String} as input variables, returns a variable \textit{R} typed \textit{Boolean} to indicate customer successful login. The required resource is defined in \textit{RDec}. In which, the login state of customer are represented in variable \textit{LoginState} with basic type \textit{Boolean}. The current customer is represented as variable \textit{CurrentCustomer} with the type \textit{Customer}. This is not a basic type such as \textit{Float, Double, String, Boolean, and Date}, but is a domain concept. This complex type should contain at least attributes such as \textit{Name:String, Passwd:String and Balance:Double}, and it can be represented as a domain conceptual class of UML. Once we define all the interfaces of the system like above interface \textit{ManageAccountIF}, we can forge the domain model by UML conceptual class diagram in Fig. \ref{domain}. The conceptual class diagram describes abstract and meaningful concepts in the problem domain, and it decomposes the problem regarding individual concepts. This is an important trophy in requirement analysis. Therefore, we can tell that service refinement can help UML to get the more precisely model about the target system in at least requirement election stage. 

\section{Formal Specifications by Service Refinement}

\subsection{Service Specification} 
We have use cases, service interfaces, and domain model of the target system at this moment. For each service, we only know the name, but we need to precisely define the semantics of service, such as when does the service can be activated, what does the service do but not how to do, by describing the system state changes carried out by the service. UML diagrams cannot describe the semantics of service, but service refinement has the triple $(\alpha_m, g_m, P_m)$, that can specify the semantics of service. For example, the semantic of service \textit{deposit()} can be defined as follow:

{\small\setlength{\mathindent}{0pt}
 \[
 \begin{array}{l}
 \mathit{Service : deposit(inNewBalance : Double, \; outR : Boolean)} \\ \\
 \mathit{Spec_{deposit}} = (\alpha, g, P) \\ 
 \alpha = \{\mathit{LoginState, inNewBalance, outR, CurrentCustomer}\} \\
 g:  \mathit{LoginState = True} \\
 p:  \mathit{inNewBalance \geq 10} \\
 R:  \mathit{CurrentCustomer.Balance' =} \\ \qquad \mathit{CurrentCustomer.Balance + inNewBalance} \wedge \mathit{outR' = True}
 \end{array}
 \]
}

\noindent The customer can make a deposit into his account through service \textit{deposit()}, this service has one input parameter \textit{inNewBalance}, and one output parameter \textit{outR}. $g$ is a guard condition describing only when resource \textit{LoginState} is equal to \textit{True}, the service \textit{deposit()} can be invoked. The design $P$ main contains precondition $p$ and postcondition $R$. $p$ describes that the state of the system before the execution of the service. $R$ describe the state of system when the execution service has finished. In the specification of service \textit{deposit()}, $p$ describes the minimum deposit amount is at least 10 dollars. $R$ describes after execution of the service, the balance of current customer is equal to the origin balance plus the new deposit account, and the value of variable \textit{ourR} is \textit{True}. All the required resource are defined in $\alpha$, which contains variables \textit{LoginState, inNewBalance, outR and CurrentCustomer}. Once we specified all the semantics of services in the interface of \textit{Manage Account}, we can get \textit{Spec} of interface \textit{ManageAccountIF}:

{\small\setlength{\mathindent}{0pt}
\[ 
\begin{array}{l}
	\mathit{Spec(ManageAccountIF)} = \{ \\
	\quad \mathit{Spec_{signUp}, Spec_{logIn} , Spec_{logOut}, Spec_{checkBalance}, Spec_{deposit}} \\
	\}
\end{array}
\]
}

\noindent At this moment, the semantics of all services are specified. However, the interactions between the interfaces of system and environment (actors) are not described. We will do it in the next section.

\subsection{The Protocol of Interface}

\noindent Service refinement can specify the interactions in the protocol \textit{Prot} of interface contract. The protocol of interface is a set of valid event traces of service request and response represented as $\langle ?m_{1}(x_{1}), !m_{1}(y_{1}),\ldots, ?m_{k}(x_{k}), !m_{k}(y_{k}) \rangle$, where $?m(x)$ represents a request of service $m$ with parameter $x$. $!m(y)$ represents the response of service $m$ with parameters $y$. For example, the protocol of \textit{ManageAccount} interface is:

{\small\setlength{\mathindent}{0pt}

 \[
 \begin{array}{l}
    
 \mathit{Prot(ManageAccountIF)} = \{ \, s^* \, |  \\ \\
   
   \; s = \langle \rangle \; \vee \\ \\
      
   \; s = \langle \mathit{(?signUp(inPersonalInformation),!signUp(outR))^?\rangle}  \; \vee  \\ \\
   
   \; s = \langle \mathit{(?signUp(inPersonalInformation),!signUp(outR))^?,} \\
   \qquad \mathit{?logIn(inUserName, inPasswd), !logIn(outR) \rangle} \; \vee \\ \\
   
   \; s = \langle \mathit{(?signUp(inPersonalInformation),!signUp(outR))^?,} \\
   \qquad \mathit{?logIn(inUserName, inPasswd), !logIn(outR),} \\
   \qquad \mathit{?checkBalance(), !checkBalance(outBalance),} \\ 
   \qquad \mathit{(?deposit(inNewBalance), !deposit(outR))^*,} \\ 
   \qquad \mathit{(?logOut(inNewBalance), !logOut(outR))^? \rangle} 
   
 \end{array}
 \]
}
\noindent Like regular expression, we introduce asterisk * at the right top of the event indicates zero or more occurrences of the preceding element and the question mark ? at the right top of the event indicates zero or one occurrence of the preceding element. The sequential invoking the services $m_1, m_2, m_3$ in the interface $I$, the trace set is $trace(I) = \{\langle \rangle \vee \langle m_1 \rangle \vee \langle m_1, m_2 \rangle \vee \langle m_1, m_2, m_3 \rangle\}$. The mediate traces are omit because the limit space of paper, we only show the initial trace $\langle \rangle$ and final trace $\langle m_1, m_2, m_3 \rangle$ in the following protocols. The protocol describes three main stories of interface \textit{ManageAccount}: 1) In the initial state of the interface, no service is invoked yet, therefore, the trace of interface is empty. 2) If the users are the new customers, they must request service \textit{signUp()} to open accounts in the system. 3) If they successfully register accounts or already have accounts in the system, they can call service \textit{login()} to log in the system, check their balance of accounts, deposit into their accounts. Then they may log out of the system.

\begin{figure}[!htb]
\begin{center}
  \includegraphics[width=0.40\textwidth]{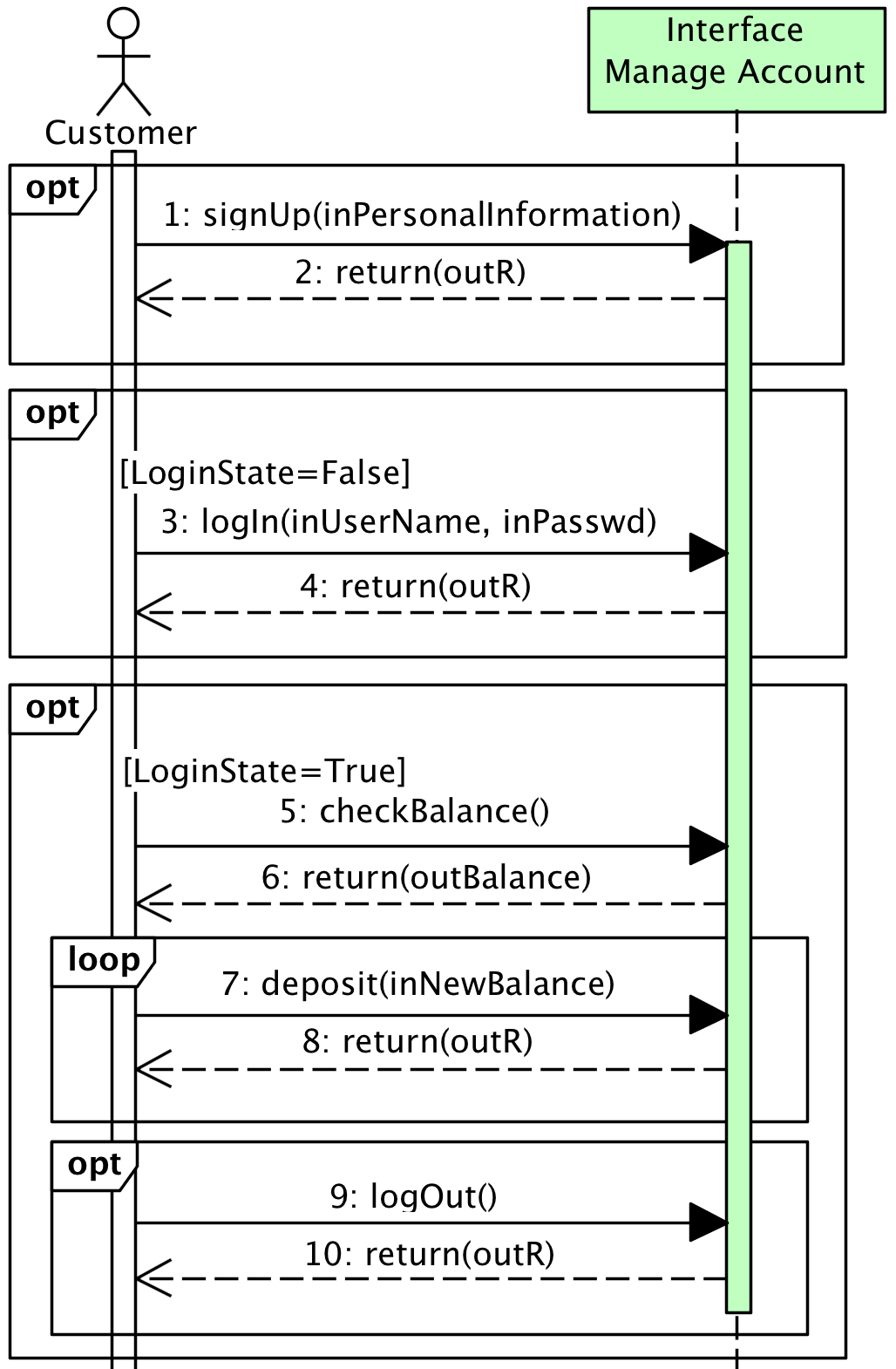}
  \caption{The Basic Event Flow of \textit{ManageAccount}}
\label{protocol}
\end{center}
\end{figure}

\subsection{System Sequence Diagram for Interface Protocol}

The protocol of interface can define the interactions of use case as above, but it is not easy to understand for the end-users, even for the developers. UML provides activity diagram and system sequence diagram, which can describe event flow in a clear way. For example, the protocol of \textit{ManageAccount} represented as system sequence diagram is in Fig. \ref{protocol}. Like using the question mark ? in the protocol to describe the optional of the event, system sequence diagram can take combined fragment \textit{opt} to describe the same situation. Combined fragment \textit{loop} can describes same loop situation as * asterisk.  Furthermore, system sequence diagram can shows the system interface and the environment actor, and the request event and response event between actor and interface. \textbf{In this point, UML can help service refinement to describe the protocol of the interface.}

%




\subsection{The Contract of Interface}
Until now, we specified the functional requirements in interfaces of system \textit{I}, the specification of services \textit{Spec} including guard condition for activating service, the states of the system before and after execution of service, and the interactions protocol of interface \textit{Prot} between the system and the environment. However, we still lack the information about the initial state of the interface. In the theory of service refinement,  the contract of interface specifies the initial state of the resources. For example, the interface \textit{ManageAccount} contains resource in \textit{RDec}:

{\small\setlength{\mathindent}{0pt}
    \[
    \begin{array}{l}
    
           \mathit{ManageAccountIF.RDec} = \{ \\
          \quad \mathit{UserDB : Customer^*}, \; \mathit{CurrentCustomer : Customer}, \\
          \quad \mathit{TempCustomer : Customer}, \; \mathit{LoginState : Boolean} \\
          \}
    \end{array}
    \]}

\noindent The initial operations of those resources are described as:

 {\small\setlength{\mathindent}{0pt}
 \[
 \begin{array}{l}
 \mathit{Init(ManageAccountIF) = True \; \vdash \; (}\\
   \quad \mathit{UserDB' = CustomerDatabase \; \wedge \; CurrentCustomer' = Null \; \wedge} \\ 
   \quad \mathit{TempCustomer' = Null \; \wedge \; LoginState' = False \;\wedge \; \neg wait'} \\
   )
 \end{array}
 \]
 }

\noindent After system initialization, the resource \textit{UserDB} contains all the customers in the database, \textit{CurrentCustomer} is a reference initialized as \textit{Null}, that means this resource is not refer to any variable. \textit{TempCustomer} is \textit{Null}, the boolean variable \textit{LoginState} is \textit{False}. The variable \textit{wait} is False, which represents the system in the stable state, and system ready for receiving service requests from environments. After specifying the initial state of interface, we can derive the contract of the interface. For example, the contract of interface \textit{ManageAccount} is specified as:

{\small\setlength{\mathindent}{0pt}
\[
\begin{array}{l}
	\mathit{Ctr(ManageAccount)} = (\\
	\quad \mathit{ManageAccountIF, \; Init(ManageAccountIF)}, \\
	\quad \mathit{Spec(ManageAccountIF), \; Prot(ManageAccountIF)} \\
	)
\end{array}
\]
}

\subsubsection{Failures}

The contract of the interface contains the specifications of resources, services, the initial state of the interface, and protocol of the interface. However, the protocol only provides the valid interactions between system and the environment. We not only need to know what the interfaces of the system can do but also what they refuse to do. Service refinement can specify the refusal scenarios by modeling \textit{Failures(Ctr)} of the contract. A failure is a pair $(s, X)$ where $s$ is the traces of interactions between environments and the system, and $X$ denote the refutation set of the services of the contract after trace $s$. \textit{Failures(Ctr)} is the set of all \textit{Ctr} failures, which defined as:

{\small\setlength{\mathindent}{0pt}
           \[
           \begin{array}{l}
           	\mathit{Failures(Ctr)=_{df}} \\ \\
           	\left \{
           	\begin{array}{l}
           		(\langle \rangle, X \}) \; | \\
           		\exists v' \cdot \mathit{Init[true,false,true,false/ok,wait,ok',wait']} \; \wedge \\
           		\forall ?m \in X \cdot \neg g_{m}[v'/v]
           	\end{array}
           	\right \}  \cup 
           \end{array}
           \]

}

\noindent This fragment of failure describes the refusal requests of services because the guards of service do not hold in the initial state of interface. For the contract of \textit{ManageAccount}, the variable \textit{LoginState} is equal to \textit{False} in the initial state. In Fig.\ref{protocol} of workflow about usecase \textit{ManageAccount}, the services \textit{checkBalance()}, \textit{deposit()}, and \textit{logOut()} can be activated only if $\mathit{LoginState = True}$. That means those services refuse to response environments at the initial state of system interface. Formally, after initialization with empty trace $\langle\rangle$, the refusal service of interface \textit{ManageAccount} are:

 {\small\setlength{\mathindent}{0pt}\[\mathit{X = \{?checkBalance(),?deposit(inNewBalance),?logOut()}\}\]}

 
\noindent The next fragment definition of contract failures are:

{\small\setlength{\mathindent}{0pt}
          \[
           \begin{array}{l}
           \left \{
           \begin{array}{l}
           (\langle ?m_{1}(x_{1}), !m_{1}(y_{1}), \dots, ?m_{k}(x_{k}), !m_{k}(y_{k})\rangle, X) \; | \\
           \exists v' \cdot (Init;\dots;(g_{m_{k}}\verb|&|P_{m_{k}})[x_{k},y_{k}/x,y]) \\
           \verb|[|true,false,true,false/ok,wait,ok',wait']  \wedge \\
           \forall ?m \in X \cdot \neg g_{m}[v'/v]
           \end{array}
           \right \}  \; \cup \\
		   \end{array}
		   \]
}

\noindent This failure set describes after execution of a sequence request events, the guards of services became false. The postcondition of \textit{login} service shows the variable \textit{LoginState' = True} after successfully logging the system. That means when the customers have already logged in the system, they can not log in again because the guard condition of \textit{login} service cannot be satisfied at this moment. Formally, after the traces of login service

	{\small\setlength{\mathindent}{0pt}
	\[
	\begin{array}{l}
		\mathit{\langle (?signUp(inPersonalInformation),!signUp(outR))^?,} \\ 
		\; \; \mathit{?login(inUserName, inPasswd), !login(outR)} \rangle
	\end{array}
	\]}
	
\noindent The refusal service are {\small$\mathit{X=\{ ?login(inUserName, inPasswd)\}}$}. After the traces of \textit{logout} service,
	{\small\setlength{\mathindent}{0pt}
	\[
	\begin{array}{l}
		\mathit{\langle (?signUp(inPersonalInformation),!signUp(outR))^?,} \\ 
		\; \; \mathit{?login(inUserName, inPasswd), !login(outR)} \\
		\; \; \mathit{?checkBalance(), !checkBalance(outBalance)} \\
		\; \; \mathit{(?deposit(inNewBalance), !deposit(outR))^*} \\
		\; \; \mathit{?logOut(), !logOut(outR)}
		\rangle
	\end{array}
	\]}
\noindent the variable \textit{LoginState} become \textit{False}. Thus, the refusal services are {\small$\mathit{X = \{ ?checkBalance(), ?deposit(), ?logOut() \}}$}. 

The next fragment of failures describe for every request of service, the system must have response to the environment, even when the service is in the fault. The formal description is:

{\small\setlength{\mathindent}{0pt}
          \[
           \begin{array}{l}
           \left \{
           \begin{array}{l}
           (\langle ?m_{1}(x_{1}), !m_{1}(y_{1}), \dots, ?m_{k}(x_{k}))\rangle, \; X) \; | \\
           \exists v' \cdot (Init;\dots;(g_{m_{k}}\verb|&|P_{m_{k}})[x_{k}/x]) \\
           \verb|[|true,false,true,false/ok,wait,ok',wait']  \wedge !m_{k} \not \in X
           \end{array}
           \right \}   \cup \\
		   \end{array}
		   \]
}

\noindent In Fig.\ref{protocol} of the event flow of contract \textit{ManageAccount}, every request event of service has a corresponding response event. Therefore, the system does not refuse to response for any request. 

\subsubsection{Deadlock}

The next fragment of failures describe the requests of services leading the system into a waiting state. The formal description is:

{\small\setlength{\mathindent}{0pt}
          \[
           \begin{array}{l}

           \left \{
           \begin{array}{l}
           (\langle ?m_{1}(x_{1}), !m_{1}(y_{1}), \dots, ?m_{k}(x_{k}))\rangle, X) \; | \\
           \exists v' \cdot (Init;\dots;\\
           (g_{m_{k-1}}\verb|&|P_{m_{k-1}})[x_{k-1},y_{k-1}/x,y])[true,false/ok',wait']; \\
           (g_{m_{k}}\verb|&|P_{m_{k}})[x_{k}/x])[true,false,true,true/ok,wait,ok',wait']
           \end{array}
           \right \} \; \cup 
		   \end{array}
		   \]
}

\noindent The \textit{deposit} service will lead system into a waiting state. For example, when a customer uses his credit card to add the balance to his account. The system will wait for the response from the third-party payment service. Usually, the system only need to wait seconds before getting the reply of the payment service, and then set the variable \textit{wait} to \textit{False} after execution the \textit{deposit} service. However, the third-party payment service may be ignoring the payment request or not receive the request because of the fault of the internet and too many requests beyond his processing ability. If the requirements are not taken the exception cases into account, that will lead a deadlock. In that situation, the system will not give any response to any the request of the services from the environments. The failure of this case can specify as, after the execution services of traces 

	{\small\setlength{\mathindent}{0pt}
	\[
	\begin{array}{l}
		\mathit{\langle (?signUp(inPersonalInformation),!signUp(outR))^?,} \\ 
		\; \; \mathit{?login(inUserName, inPasswd), !login(outR)} \\
		\; \; \mathit{?checkBalance(), !checkBalance(outBalance)} \\
		\; \; \mathit{?deposit(inNewBalance)} \rangle
	\end{array}
	\]}

\noindent The system refuses to response any requests of public services $\mathit{X = \{?m \; | \; m \in public(ManageAccountIF.SDec)\}}$. There is a terrible property of the system, and we must refine the contract to the deadlock-free requirements before making any implementation. The simple idea to handle this situation is designed an internal service \textit{repeatInvolkingPayment()} to periodically such as 30s send the request to the third-party payment service until it is successfully getting the response. Note that the \textit{repeatInvokingPayment()} must be a deadlock-free service. It has a inside timer to count the waiting time, once it does not can response from the third-part payment service till the maximum waiting time, it will force \textit{repeatInvokingPayment()} to return a result. It is the useful strategy against the fault of the internet such as lost package and denies offer service of the third-party system. 

\subsubsection{Livelock}

The interface in class diagram can also have the ability to describe the internal service. The services in UML interface diagram of Fig.\ref{interface} are public with the prefix mark "+", the internal service is private with the prefix mark "-". However, this internal service will make the system even worse. When third-party service is dead forever, \textit{repeatInvolkingPayment()} service will never stop to send the request. That will make the system only make this internal service running forever, but not make any useful response to the environment. That is so-called livelock of the system, or the system is in the diverged state. Once the system stuck into the livelock, the system will refuse to responses from the environment. That is the last situation specified in the failures of the contract. Formally, the fragment of failures about divergences of the contract are defined as the set of the pair {\setlength{\mathindent}{0pt}$\; \{ \; (s, X)\; | \; s \in Divergences(Ctr) \; \}$}. Divergences shows the request event of service $\mathit{?m_{k}(x_k)}$ will lead system into a unstable state where $ok'$ is \textit{false}. In divergence state, the internal services are infinitely invoked, that cannot affect system into a stable state. In our case, the internal service \textit{repeatInvolkingPayment()} will lead system into a divergences state when the third-party service is forever unavailable. Formally, the divergence of interface \textit{ManageAccount} is:
        	
  	  	{\small\setlength{\mathindent}{0pt}     
        \[
        \begin{array}{l}
           \mathit{Divergences(ManageAccountIF)=_{df}} \\
            	 \; \; \mathit{\langle (?signUp(inPersonalInformation),!signUp(outR))^?,} \\ 
				 \; \; \mathit{?login(inUserName, inPasswd), !login(outR),} \\
				 \; \; \mathit{?checkBalance(), !checkBalance(outBalance),} \\
			     \; \; \mathit{?deposit(inNewBalance),}  \mathit{?repeatInvokingPayment()^*} \rangle\; | \; 
            (Init;\dots; \\
           		\; \; (\mathit{g_{checkBalance}}\verb|&|\;\mathit{P_{checkBalance})}[\mathit{outBalance}/y] \mathit{[true,false/ok',wait']}) \\
           		\; \; (\mathit{g_{deposit}}\verb|&|\mathit{P_{deposit}})[\mathit{inNewBalance}/x] \mathit{[true,false/ok,ok']}) \rangle
        \end{array}
        \]  
    }       
		   
\noindent Once we specified divergence of \textit{ManageAccountIF}, the last failures fragment of interface manage account are \( \mathit{\{(s, public(ManageAccountIF.SDec))\; | \; s \in Divergences(Ctr) \}}\). In the divergence state, the interface \textit{ManageAccount} will refuse to response any services of the interface.

In this section, we show how to use service refinement to specify all possible valid traces from the initial state of the interface with the refusal service sets.  That precisely defines the semantics of the interface. We also see that UML diagrams can help to give a clue for modeling each case. UML diagrams and service refinement are complementary for each other. Although we get all the possible traces and refusal sets, that may contain the deadlock and livelock. The requirements model must meet at least functional correctness with the deadlock-free and livelock-free before implementation. 

\section{Refinement and Verification}
In this section, we will show how to refine the contracts of interfacess to the deadlock-free and livelock-free contracts through theory of service refinement, and use model checking tool to confirm the refinements.

\subsection{Contract Refinement}
The main refinement strategy is to add more control strategies and resources to make refusal set is a proper subset of all the services set $\mathit{X \subset ManageAccountIF.SDec}$ from $\mathit{X = ManageAccountIF.SDec}$. As we mentioned in the previous section, failures set of contract \textit{ManageAccount} has two refusal all service pairs with the traces while executing of deposit service, first one is in the deadlock state because the fault of request to the third-party payment service, the second one is in the livelock (divergence) state when the infinitely repeat to request the third-party payment service by the internal \textit{repeatInvokingPayment()} service. We specify the original contract of \textit{ManageAccount} as $\mathit{Ctr_1}$, and the added internal service \textit{repeatInvokingPayment()} contract as $\mathit{Ctr_2}$. Although contract $\mathit{Ctr_2}$ is deadlock-free contract, $\mathit{Ctr_2}$ does not refine $\mathit{Ctr_1}$ because they violate the refinement definition. The contract $\mathit{Ctr_2}$ is more divergence than $\mathit{Ctr_1}$: $\textbf{Divergences}(Ctr_{2})\supseteq \textbf{Divergences}(Ctr_{1})$. We continue to improve the $\mathit{Ctr_2}$ to livelock-free contract by adding a variable \textit{MaxRepeatedTimes} typed \textit{Integer} to the \textit{RDec} of interface \textit{ManageAccount}, and add an control strategy, that once invoking \textit{repeatInvokingPayment()} reach $\mathit{MaxRepeatedTimes}$ times (e.g, 3 times), it will return the result $\mathit{!deposit(false)}$, and set $\mathit{wait'}$ as $\mathit{false}$. At that moment, the refusal service set $X$ of the contract $\mathit{Ctr_3}$ will not contain all public services $\mathit{X \not= public(ManageAccount.SDec)}$ after the traces:

	{\small\setlength{\mathindent}{0pt}
	\[
	\begin{array}{l}
		\mathit{\langle (?signUp(inPersonalInformation),!signUp(outR))^?,} \\ 
		\; \; \mathit{?login(inUserName, inPasswd), !login(outR)} \\
		\; \; \mathit{?checkBalance(), !checkBalance(outBalance), } \\
		\; \; \mathit{?deposit(inNewBalance), ?repeatInvokingPayment()^{MaxRepeatedTimes}}, \\ 
		\; \; \mathit{!deposit(false)} \rangle 
	\end{array}
	\]}

\noindent After adding the variable \textit{MaxRepeatedTimes} and maximum trying strategy to contract $\mathit{Ctr_3}$ of interface \textit{ManageAccount}, while request service \textit{?deposit(inNewBalance)} service, the environment eventually receives the response \textit{!deposit(outR)} after no more than \textit{MaxRepeatedTimes} times invoking the internal service \textit{repeatInvokingPayment()}. $\mathit{Ctr_3}$ contract will not make system into a divergence state. Therefore, the contract $\mathit{Ctr_3}$ is a deadlock-free and livelock-free contract, and $\mathit{Ctr_3}$ refines $Ctr_1$ because after hiding the private service set $PM_3$ = \textit{\{repeatInvokingPayment()\}}, the failures and divergences of $\mathit{Ctr_3}$ and $Ctr_1$ that holds: 

\begin{description}
   \item [{(1)}] \( \textbf{Divergences}(Ctr_{1})\supseteq
        \textbf{Divergences}(Ctr_{3}/PM_3)\)
   \item [{(2)}] \( \textbf{Failures}(Ctr_{1})\supseteq
        \textbf{Failures}(Ctr_{3}/PM_3)\)
\end{description}

\noindent Furthermore, the contract $\mathit{Ctr_3}$ is a consistency contract, it will never enter deadlock and livelock states if environment follows the protocol of the contract. Formally, the $\mathit{Prot}$ of $\mathit{Ctr_3}$ is:

{\small\setlength{\mathindent}{0pt}

 \[
 \begin{array}{l}
    
 \mathit{Prot(Ctr_3)} = \{ \, s^* \, | \\ \\
 
   \; s = \langle \rangle \; \vee \\ \\
      
   \; s = \langle \mathit{(?signUp(inPersonalInformation),!signUp(outR))^?\rangle}  \; \vee  \\ \\
   
   \; s = \langle \mathit{(?signUp(inPersonalInformation),!signUp(outR))^?,} \\
   \qquad \mathit{?logIn(inUserName, inPasswd), !logIn(outR) \rangle} \; \vee \\ \\
   
   \; s = \langle \mathit{(?signUp(inPersonalInformation),!signUp(outR))^?,} \\
   \qquad \mathit{?logIn(inUserName, inPasswd), !logIn(outR),} \\
   \qquad \mathit{?checkBalance(), !checkBalance(outBalance),} \\ 
   \qquad \mathit{(?deposit(inNewBalance),}\\  
   \qquad \mathit{?repeatInvokingPayment()^{0..MaxRepeatedTimes}}, \\
   \qquad \mathit{!deposit(outR))^*,}  
   \mathit{(?logOut(), !logOut(outR))^? \rangle}
 \end{array}
 \]
}

\noindent The only difference between the protocols of $\mathit{Ctr_3}$ and $\mathit{Ctr_1}$ is in the last fragment 
$\mathit{Ctr_3}$ contains the request events of \(\mathit{?repeatInvokingPayment()^{0..MaxRepeatedTimes}}\) where the number 0..MaxRepeatedTimes represents the repeat times of the event. The $\mathit{Failures}$ of $\mathit{Ctr_3}$ is:

{\small\setlength{\mathindent}{0pt}
           \[
           \begin{array}{l}
           	\mathit{Failures(Ctr_3)= \{} \\ \\
           	
           	\mathit{(\langle \rangle, X = \{?checkBalance(),?deposit(inNewBalance),?logOut())}, \\ \\

           	\mathit{(\langle ?signUp(inPersonalInformation),!signUp(outR)) \rangle,} \\ 
           	\quad  \mathit{X = \{?checkBalance(),?deposit(inNewBalance),?logOut(),} \\ 
           	\qquad \mathit{?signUp(inPersonalInformation))}\}), \\ \\
            
           	\mathit{(\langle (?signUp(inPersonalInformation),!signUp(outR))^?,} \\ 
			\qquad \mathit{?logIn(inUserName, inPasswd), !logIn(outR)} \rangle, \\
			\quad \mathit{X=\{ ?logIn(inUserName, inPasswd)\}}), \\\\

			\mathit{(\langle (?signUp(inPersonalInformation),!signUp(outR))^?,} \\
   			\qquad \mathit{?logIn(inUserName, inPasswd), !logIn(outR),} \\
   			\qquad \mathit{?checkBalance(), !checkBalance(outBalance),} \\ 
   			\qquad \mathit{(?deposit(inNewBalance),} \\
   			\qquad \mathit{?repeatInvokingPayment()^{MaxRepeatedTimes}}, \\

			\quad \mathit{X = \{ ?checkBalance(), ?deposit(), ?logOut(),} \\ 
			\qquad \mathit{?repeatInvokingPayment() \}}), \\ \\
			
			\mathit{(\langle (?signUp(inPersonalInformation),!signUp(outR))^?,} \\
   			\qquad \mathit{?logIn(inUserName, inPasswd), !logIn(outR),} \\
   			\qquad \mathit{?checkBalance(), !checkBalance(outBalance),} \\ 
   			\qquad \mathit{(?deposit(inNewBalance),} \\
   			\qquad \mathit{?repeatInvokingPayment()^{0..MaxRepeatedTimes}}, \\
   			\qquad \mathit{!deposit(outR))^*,}  
   			\mathit{(?logOut(inNewBalance), !logOut(outR))^? \rangle,} \\

			\; \; \mathit{X = \{ ?checkBalance(), ?deposit(), ?logOut() \}}),

           \\	
		   \}
           \end{array}
           \]
}

\noindent The difference between the failures of $\mathit{Ctr_3}$ and $\mathit{Ctr_1}$ is that $\mathit{Failures(Ctr_3)}$ does not contains the deadlock and livelock pairs in $\mathit{Failures(Ctr_1)}$. Therefore, the contract $\mathit{Ctr_3}$ is  consistency contract, which holds:

\begin{description}
    \item [{(1)}] \(  \mathit{\forall tr \in Prot(Ctr_3) \cdot (\exists s \in traces(Ctr_3) \cdot s \downarrow \{?\} = tr)}\)
	\item [{(2)}] \( \mathit{\forall tr \in Prot \; \forall(s, X) \in failures(Ctr_3) \cdot (s \downarrow \{?\} } \preceq tr \Rightarrow \\ X \not = \{ ?m,!m \; | \; m \in \mathit{public(MDec_3))} \cup \textit{private(MDec)}\}\) \(
\wedge s \downarrow \{?\} \not = \{ \langle ?m^* \rangle \; | \; ?m \in \textit{private(MDec) \}} 
\)
\end{description}

\noindent First condition represents, for every trace of the protocol, failures have the corresponding specification to describe the refusal sets of the services requests. Second condition shows, for every the trace of the protocol, the system must have at least one service can be invoked, and the subsequences trace of $s$ does not include infinitely invoking the private services. That means the consistency contract must be deadlock-free and livelock-free.

\subsection{Contract Verification}
\begin{figure}[!htb]
\begin{center}
  \includegraphics[width=0.3\textwidth]{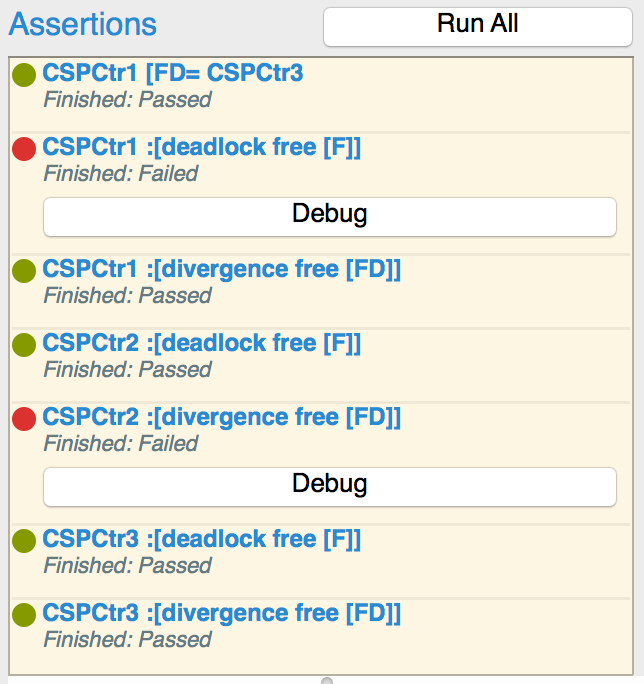}
  \caption{FDR Verfication}
\label{verication}
\end{center}
\end{figure}

The semantics of service contract is defined by the failure and divergence models from CSP. To prove the validity of our work, we model and verify service contracts \textit{$Ctr_1$}, \textit{$Ctr_2$}, and \textit{$Ctr_3$} in CSP model checking tool FDR \cite{gibson2014fdr3}. The verification result is shown in Fig. \ref{verication}. 

\begin{figure}[!htb]
\begin{center}
  \includegraphics[width=0.5\textwidth]{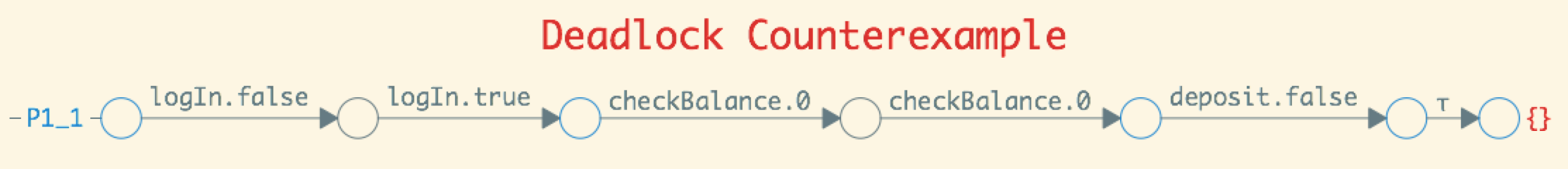}
  \caption{Deadlock Counter Example of Ctr1}
\label{deadlock}
\end{center}
\end{figure}

\noindent Service contract \textit{$Ctr_1$} is divergence-free but not deadlock-free, because the \textit{?deposit()} can make system into a deadlock state for waiting the return result from a third-part payment service, that is presented as an counter example in Fig. \ref{deadlock}.

\begin{figure}[!htb]
\begin{center}
  \includegraphics[width=0.5\textwidth]{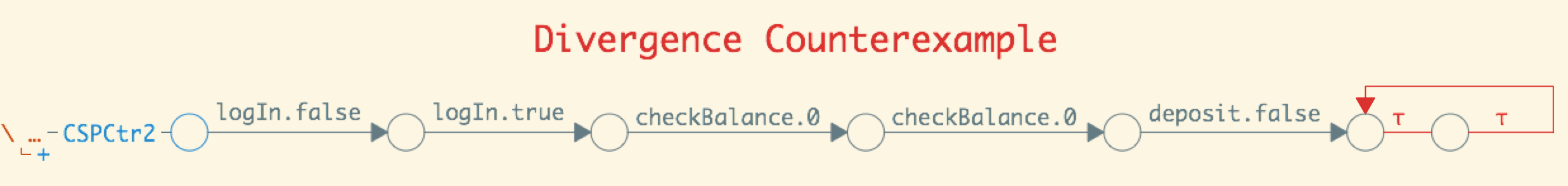}
  \caption{Divergence Counter Example of Ctr2}
\label{divergence}
\end{center}
\end{figure}

\noindent After introducing the private service \textit{repeatInvokingPayment()} in contract \textit{$Ctr_2$}, the contract is not longer deadlock-free, because the service \textit{deposit()} periodically invoke this private service \textit{repeatInvokingPayment()}  until successfully getting the response. However, the contract \textit{$Ctr_2$} will stuck in a livelock state because infinitely invoking the \textit{repeatInvokingPayment()} service once the third-part payment service is permanently shutdown. This divergence counter example is shown in Fig. \ref{divergence}.

To deal with livelock, we introduce a control strategy with maximum repeat limitation to the private service \textit{repeatInvokingPayment()} as contract \textit{$Ctr_3$}. Which will make the system only try \textit{MaxRepeatedTime} times, then make return a false result. As shown in Fig. \ref{verication}, contract \textit{$Ctr_3$} is deadlock-free and livelock-free, and the refinement verification result shows contract \textit{$Ctr_3$} is a refinement from contract \textit{$Ctr_1$}. In short, we get the same verification result from the model checking tool. That ensure the validity of our work in this paper.

Note that we did not show the FDR model in this paper, you can reach it from the GitHub gist\footnote{\url{https://gist.github.com/yylonly}}.

\section{Related Work} 
All the related work about service refinement are listed. The paper\cite{li2010denotational} proposes a denotational semantical model to service orchestration language with the service refinement, and it can determine whether service orchestration satisfies its specification. The paper\cite{wang2009method} based on service refinement proposed a formal model for web service interfaces, mismatches detecting among multiple web services. The papers \cite{li2009towards, li2014formala} propose a formal model to specify and analyze the behavior and robustness of service mashups under an unstable environment. The paper \cite{zhao2010probabilistic} present a computable probabilistic model of the survivable system based on the service refinement. The paper \cite{zhang2009web} used service refinement for providing a mathematical model for WSDL 2.0. The paper \cite{li2010promoting} proposes a concept of promoting models to obtain refinements with support from cooperating models. The papers \cite{liu2006reactive} and \cite{herold2008cocome} are the case studies of this theory but does not include the contract refinements, and system properties are checking about divergence, failure, and deadlock.  To best our knowledge, all the related works do not touch the integrate service refinement with UML requirement analysis to support use case contract refinement and consistent checking and do not demonstrate the power of UML to help to elicit the start-up requirements for service refinement. 

Other formal methods are also considered to integrate with UML for system modeling and verification. The paper \cite{evans1998uml} motivates an approach to formalizing UML in which formal specification techniques are used to gain insight into the semantics of UML notations and diagrams. A small example is presented through Z notation to verify whether one class diagram is a valid deduction of another. Another work \cite{borges2007integrating} integrate UML class diagrams and OhCircus by written UML elements in terms of OhCircus constructs. UML-B \cite{snook2006uml,soton264926} provides an UML front-end for B methods and Event-B, which provides a formally precise variant of UML to support model refinement. The paper \cite{beato2005uml} provides an tools for automatically verifying UML model after transforming the active behaviour from UML activity diagram and class diagram into SMV. In short, the related works focus on verifying the design model (class diagram) of the system specified by UML. Our approach make the verification more early in the requirements model, because 1) the problems in requirements model can be passed to the design model, and 2) to verify a design model, we not only need to specify a requirements model first but also need efforts to specify a design model. To be safety and make less efforts, our approach verified the system in the stage of software engineering process as earlier as possible. This is mainly novelty idea our approach. \textbf{Note that} we only focus on specifying and verifying the requirements model specified by UML, which includes use case diagram, interface diagram, conceptual class diagram and system sequence diagram (activity diagram). Other diagram for system design such as class diagram, sequence diagram (collaboration diagram), component diagram and state diagram are not included.

\section{Conclusions and Future Work}
Because of the limited space, we only show the contract of the interface \textit{ManageAccount} in the paper. Other consistency contracts of interfaces can be specified and then refined in the same approach. This paper shows the both advantages of UML and service refinement in the procedures of the requirement analysis. As usual, the developers and researcher use UML and theory of formal methods like service refinement independently. This paper shows how to integrate each other to do better work. The developers can use their favourite UML tools along with service refinement to elicit the consistency requirements with reduced the number of bugs at the early stage of software engineering. The formal methods researchers can use UML diagrams to elicit the start-up requirements for next round specifications by their formal approach. Hopefully, this paper can make an attention for both the developers and researchers in the software engineering to do more work on integrating the software engineering industry approach with the formal approach.

This paper only touches the procedures of in requirement analysis. We can eventually get the consistency contracts of interface for all the use cases of the system. The contracts can be further designed and then implemented in the object-oriented, component-based, and service-oriented approaches. Furthermore, we consider to integrate the formal approach with our prototype generation tool RMCode to support agile requirement election and user-level requirement validation.

\section*{Acknowledgement}

This work was supported by the Macau Science and Technology Development Fund (FDCT) with No. 103/2015/A3 and the National Natural Science Foundation of China (NSFC) with grant No. 61562011 and 61672435.
\newpage
\noindent

\bibliographystyle{ieicetr}
\bibliography{Component}

\begin{thebibliography}{10}

\bibitem{sommerville2015software}
I.~Sommerville, ``Software engineering. international computer science
  series,'' ed: Addison Wesley, 2015.

\bibitem{larman2012applying}
C.~Larman, Applying UML and Patterns: An Introduction to Object Oriented
  Analysis and Design and Interative Development, Pearson Education India,
  2012.

\bibitem{saeeiab2000method}
S.~Saeeiab and M.~Saeki, ``Method integration with formal description
  techniques,'' IEICE transactions on Information and Systems, vol.83, no.4,
  pp.616--626, 2000.

\bibitem{oda2017formal}
T.~ODA, K.~ARAKI, and P.G. LARSEN, ``A formal modeling tool for exploratory
  modeling in software development,'' IEICE Transactions on Information and
  Systems, vol.100, no.6, pp.1210--1217, 2017.

\bibitem{fan2006constraint}
C.F. Fan and C.Y. Cheng, ``Constraint-based software specifications and
  verification using uml,'' IEICE transactions on information and systems,
  vol.89, no.6, pp.1914--1922, 2006.

\bibitem{hjfo8}
J.~He, ``Service refinement,'' Science in China Series F: Information Sciences,
  vol.51, no.6, pp.661--682, 2008.

\bibitem{hoare1998unifying}
C.A.R. Hoare and H.~Jifeng, Unifying theories of programming, Prentice Hall
  Englewood Cliffs, 1998.

\bibitem{csp}
C.A.R. Hoare, Communicating Sequential Processes, Prentice-Hall, 1985.

\bibitem{6960112}
W.~Yu, C.G. Yan, Z.~Ding, C.~Jiang, and M.~Zhou, ``Modeling and verification of
  online shopping business processes by considering malicious behavior
  patterns,'' IEEE Transactions on Automation Science and Engineering, vol.13,
  no.2, pp.647--662, April\ 2016.

\bibitem{hopson2009online}
D.B. Hopson and K.S. Keys, ``Online shopping system,'' Sept.~15\ 2009.
\newblock US Patent 7,590,567.

\bibitem{gibson2014fdr3}
T.~Gibson-Robinson, P.~Armstrong, A.~Boulgakov, and A.W. Roscoe, ``Fdr3—a
  modern refinement checker for csp,'' International Conference on Tools and
  Algorithms for the Construction and Analysis of Systems, pp.187--201,
  Springer, 2014.

\bibitem{li2010denotational}
Q.~Li, H.~Zhu, and J.~He, ``A denotational semantical model for orc language,''
  Theoretical Aspects of Computing--ICTAC 2010, pp.106--120, 2010.

\bibitem{wang2009method}
S.~Wang, G.~Zhang, X.~Zhang, and Y.~Yang, ``A method for detecting behavioral
  mismatching web services,'' Web Information Systems and Applications
  Conference, 2009. WISA 2009. Sixth, pp.116--121, IEEE, 2009.

\bibitem{li2009towards}
Q.~Li and H.~Zhu, ``Towards specification and refinement of contracts with
  environment changes,'' Software Engineering Workshop (SEW), 2009 33rd Annual
  IEEE, pp.61--68, IEEE, 2009.

\bibitem{li2014formala}
Q.~Li, J.~Shi, and H.~Zhu, ``A formal framework for service mashups with
  dynamic service selection,'' Innovations in Systems and Software Engineering,
  vol.10, no.3, pp.219--234, 2014.

\bibitem{zhao2010probabilistic}
Y.~Zhao, Y.~Huang, J.~Li, and H.~Zhu, ``Probabilistic model of system
  survivability,'' Theoretical Aspects of Software Engineering (TASE), 2010 4th
  IEEE International Symposium on, pp.193--200, IEEE, 2010.

\bibitem{zhang2009web}
A.~Zhang and X.~Xie, ``Web services semantic model system,''
  Anti-counterfeiting, Security, and Identification in Communication, 2009.
  ASID 2009. 3rd International Conference on, pp.592--595, IEEE, 2009.

\bibitem{li2010promoting}
Q.~Li, Y.~Zhao, X.~Wu, and S.~Liu, ``Promoting models.,'' UTP, pp.234--252,
  Springer, 2010.

\bibitem{liu2006reactive}
J.~Liu and J.~He, ``Reactive component based service-oriented design-a case
  study,'' Engineering of Complex Computer Systems, 2006. ICECCS 2006. 11th
  IEEE International Conference on, pp.10--pp, IEEE, 2006.

\bibitem{herold2008cocome}
S.~Herold, H.~Klus, Y.~Welsch, C.~Deiters, A.~Rausch, R.~Reussner, K.~Krogmann,
  H.~Koziolek, R.~Mirandola, B.~Hummel, {\em et~al.}, ``Cocome-the common
  component modeling example,'' The Common Component Modeling Example,
  pp.16--53, 2008.

\bibitem{evans1998uml}
A.~Evans, R.~France, K.~Lano, and B.~Rumpe, ``The uml as a formal modeling
  notation,'' International Conference on the Unified Modeling Language,
  pp.336--348, Springer, 1998.

\bibitem{borges2007integrating}
R.M. Borges and A.C. Mota, ``Integrating uml and formal methods,'' Electronic
  Notes in Theoretical Computer Science, vol.184, pp.97--112, 2007.

\bibitem{snook2006uml}
C.~Snook and M.~Butler, ``Uml-b: Formal modeling and design aided by uml,'' ACM
  Transactions on Software Engineering and Methodology (TOSEM), vol.15, no.1,
  pp.92--122, 2006.

\bibitem{soton264926}
C.~Snook and M.~Butler, ``Uml-b and event-b: an integration of languages and
  tools,'' February\ 2008.

\bibitem{beato2005uml}
M.E. Beato, M.~Barrio-Sol{\'o}rzano, C.E. Cuesta, and P.~de~la Fuente, ``Uml
  automatic verification tool with formal methods,'' Electronic Notes in
  Theoretical Computer Science, vol.127, no.4, pp.3--16, 2005.

\end{thebibliography}

\vspace{.3cm}
\profile[yyl.pdf]{Yilong Yang}{received the B.S. degree in Computer Science from China University of Mining and Technology, China in 2010. The M.S. degree from the Guizhou university, China in 2013, and he was a follow in United Nations University - International Institute for Software Technology, Macau. Currently, he is a PhD candidate of Software Engineering in University of Macau. His research interests include software engineering and machine learning. }
\profile[yj.pdf]{Jing Yang}{received the B.S. degree in Math from Southwestern Normal University, China in 1990. The M.S. and Ph.D. degree in Math and Computer Science from Guizhou University, China in 1993 and 2006. She was a research follow in United Nations University - International Institute for Software Technology, Macau from 2003 to 2004 and Hong Kong University of Science and Technology in 2007. Currently, she is a  professor of College of Computer Science and Technology in Guizhou University. Her research interests cover mathematical logic and formal method. }
\profile[fstxsl]{Xiaoshan Li}{got his Ph.D. degree in the Institute of Software, the Chinese Academy of Sciences, Beijing in 1994. Currently, he is an Associate Professor in Department of Computer and Information Science of the Faculty of Science and Technology (FST-UMAC). His research interests are formal methods, object-oriented software engineering with UML, real-time specification and verification, the semantics of programming language. He won the second class in the Macao Natural Science Award of Macao 2012 for the project of "rCOS Formal Model Driven for Software Development" (collaborated with Zhiming Liu).}

\end{document}